\documentclass[twocolumn,showpacs,showkeys,preprintnumbers,prl,amsmath,amssymb]{revtex4}
\usepackage[dvips]{graphicx}

\newcommand{\ab}{$\sim$ }
\newcommand{\aby}{$\sim$}

\newcommand{\be}{\begin{equation} }

\newcommand{\cmc}{cm$^{3}$ }
\newcommand{\cmcy}{cm$^{3}$}
\newcommand{\cms}{cm$^{2}$ }
\newcommand{\cmsy}{cm$^{2}$}

\newcommand{\ene}{\end{equation}}

\newcommand{\eqr}{Eq.~\ref}

\newcommand{\estary}{$E^{*}$}

\newcommand{\figr}{Fig.~\ref}

\newcommand{\hcly}{$B_{c1}$}

\newcommand{\hcu}{$B_{c2}$ }

\newcommand{\hcuy}{$B_{c2}$}

\newcommand{\itt}{$I(t)$ }
\newcommand{\itty}{$I(t)$}
\newcommand{\jj}{$j$ }
\newcommand{\jjy}{$j$}
\newcommand{\jc}{$j_{c}$ }
\newcommand{\jcy}{$j_{c}$}
\newcommand{\jd}{$j_{d}$ }

\newcommand{\jdy}{$j_{d}$}
\newcommand{\je}{$j(E)$ }

\newcommand{\lam}{$\lambda$ }
\newcommand{\lamy}{$\lambda$}

\newcommand{\lb}{\left(}

\newcommand{\lk}{$L_k$ }

\newcommand{\slco}{Sr$_{1-x}$La$_{x}$CuO$_{2}$ }

\newcommand{\mgb}{MgB$_{2}$ }

\newcommand{\rb}{\right)}
\newcommand{\rfff}{$\rho_{f}$ }

\newcommand{\rrhofy}{$\rho_{f}$}

\newcommand{\rrho}{{$\rho$} }
\newcommand{\rrhon}{$\rho_{n}$ }
\newcommand{\rrhony}{$\rho_{n}$}
\newcommand{\rrhoy}{{$\rho$}}
\newcommand{\rrhos}{{$\rho_s$} }
\newcommand{\rrhosy}{{$\rho_s$}}

\newcommand{\scv}{superconductivity }
\newcommand{\scvy}{superconductivity}

\newcommand{\scg}{superconducting }

\newcommand{\ssc}{superconductor }
\newcommand{\sscy}{superconductor}
\newcommand{\taue}{$\tau_{\varepsilon}$ }

\newcommand{\taueey}{$\tau_{ee}$}

\newcommand{\tauepy}{$\tau_{ep}$}

\newcommand{\tc}{$T_{c}$ }

\newcommand{\tcy}{$T_{c}$}

\newcommand{\vstary}{$v^{*}$}

\newcommand{\vtt}{$V(t)$ }
\newcommand{\vtty}{$V(t)$}
\newcommand{\ybco}{Y$_{1}$Ba$_{2}$Cu$_{3}$O$_{7}$ }
\newcommand{\ybcoy}{Y$_{1}$Ba$_{2}$Cu$_{3}$O$_{7}$}

\newcommand{\nccoy}{Nd$_{2-x}$Ce$_{x}$CuO$_{4}$}
\newcommand{\row}{\rightarrow }
\begin{document}

\preprint{Quantum Studies: Mathematics and Foundations 5, 111-121 (2018)}

\title{Dissipative-regime measurements as a tool for confirming and characterizing near-room-temperature superconductivity\thanks{This work was supported by the U. S. Department of Energy, Office of Science, Office of Basic Energy Sciences, under grant number DE-FG02-99ER45763.}}


\author{Charles L. Dean and Milind N. Kunchur} 

\email[Corresponding author email: ]{kunchur@sc.edu} 
\homepage{http://www.physics.sc.edu/~kunchur}


\affiliation{Department of Physics and Astronomy, University of South
Carolina, Columbia, SC 29208}



\begin{abstract}
The search for new superconducting materials approaching room temperature benefits from having a variety of testing methodologies to confirm and characterize the presence of superconductivity. Often the first signatures of new superconducting species occur incompletely and in very small volume fractions. These trace amounts may be too weak to produce an observable Meissner effect and the resistance may not go completely to zero if the percolation threshold is not met. Under these conditions, secondary behavior---such as transitions or cross overs in the temperature dependence of magnetoresistance,  magnetic irreversibility, or thermopower---are often used as indications for the presence of \scvy . 
Our group has developed a rather unique set of fast-timescale and dissipative transport measurements that can provide another tool set for confirming and characterizing suspected superconductivity. Here we provide some background for these methods and elucidate their collaborative value in the search for new superconducting materials.

\end{abstract}

\maketitle

\section{Introduction}
\label{intro}
The recent observation of 203 K \scv in H$_3$S \cite{eremets} under pressure and possible hints of \scv above room temperature in graphite \cite{esquinazi} have once again fired up interest and hope in the detection and synthesis of materials that superconduct at and above room temperatures. The potential technological impact and scientific importance of the realization of room-temperature superconductivity
needs no further explanation or emphasis. The initial detection of a new \scg species often occurs in trace amounts in inhomogeneous samples. In such cases, the classic abrupt resistance drop to zero and the appearance of clear cut Meissner flux expulsion may be luxuries that cannot be had. Even in the case of homogeneous samples that do show the above signatures, there remains the further issue of characterizing the \scg parameters. The mixed-state upper critical field \hcuy , reflective of the coherence length $\xi$, and the penetration depth \lamy, reflective of the superfluid density \rrhosy =$1/\lambda^2$, are two crucial measurements that are amongst the first to be performed. There are multiple techniques for determining such parameters, each of which has its own advantages and limitations. Our group has developed some uncommon---and in some cases  unique---experimental techniques that investigate \sscy s at ultra-short time scales, and 
under unprecedented and extreme conditions of current density \jjy , electric fields $E$, and 
power density $p=\rho j^2$ (where \rrho is the resistivity). These techniques have led to the discovery or confirmation of several novel phenomena and regimes in \sscy s, and in addition provide an alternative method to glean information on  fundamental \scg parameters, which in some cases may be hard to obtain by other methods. 
These techniques therefore bear a special relevance in collaborative work that involves the search for new \scg materials.

\section{Background}
\label{background}
\begin{figure}
 \includegraphics[width=0.85\hsize]{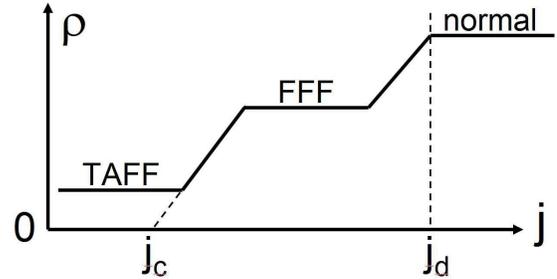} 
\caption{Progression of stages of dissipation, which alternate between Ohmic ($R$=const) and non-linear regimes.}
\label{TAFF}       
\end{figure}

In the discussion that follows, it will be assumed that the \ssc is of type II and enters the mixed state with quantized magnetic flux vortices above a lower critical magnetic field \hcly . 
As the transport current density \jj in a \ssc is steadily increased, the behavior of the \ssc goes through a progression of stages \cite{blatter} as illustrated in \figr{TAFF}. 
If not in the Meissner state and if vortices are present, there is initially some finite resistance because of flux motion. In the limit of very low driving forces, this regime can be Ohmic ($R$=const) and can be described by process such as TAFF (thermally activated flux flow) or variations thereof. As $j$ is increased, 
there is a non-linear response as vortices get depinned and commence to move more easily leading to increasing resistivity. At some point the vortices become completely depinned and flow freely. Under this condition, the resistivity stops increasing and there is a intermediate plateau of constant 
\rrhoy=\rrhofy. 
The resistivity \rfff in this free-flux-flow (FFF) state bears a relatively simple relationship, often referred to as the Bardeen-Stephen equation, to the normal-state resistivity \rrhon and the applied $B$:  
\be \rho_f \sim \rho_n B/B_{c2} \label{BS} \ene
This relationship at once ties the measurable quantity \rfff to \rrhon and \hcuy . An understanding of the normal state---the nature of the charge carriers, their concentration, and their scattering rates---is a important step in developing an understanding of the superconducting state. Therefore a knowledge of \rrho over the entire temperature range, including well below \tcy , provides an important step in elucidating the normal state. If \hcu is known, \eqr{BS} provides \rrhony ; on the other hand if \rrhon is known, \eqr{BS} provides \hcu (which may be impossible to reach at $T \ll T_c$ because of possibly prohibitively high values). Flux flow thus provides a window to the normal state and upper critical field well below \tc without the need to suppress the \scg state through very high magnetic fields \hcuy $(T \ll T_c)$. The main obstacle to this approach is that unless one is in a free-flux-flow state, presence of pinning alters \eqr{BS} in potentially complicated ways making it less reliable to relate the observed \rrho to basic parameters. Overcoming pinning requires applying a Lorentz driving force $\vec{F_L}= d(\vec{j \times \Phi_0})$ (where $d$ is the thickness) that well exceeds the pinning force $F_p$. The current density required for this is often of such a high value, that the superconductor becomes intensely dissipative. As an example, in optimally pinned \ybco (where the conventional critical current density \jc approaches the pair-breaking value \jdy ), one needs a 
$j \sim j_d >10^8$ A/\cms at low temperatures. Given that \rrhon \ab 100 $\mu\Omega$-cm for \ybcoy , this entails a power dissipation density of $p=\rho j^2 > 10^{12}$ W/\cmcy . This is of course a worst-case scenario, and the majority of \scg systems will not have such a combination of simultaneously high \rrhon and \jc (ultra high pinning); however, it is common for the dissipation densities to get above $10^{8}$ W/\cmcy, which can cause an objectionable temperature rise, if not sample destruction, if a continuous excitation is used for the transport $IV$ curve. The experimental confirmation of the Bardeen-Stephen law of \eqr{BS},  was in fact one of the motivations for us to develop the pulsed-signal techniques (described in the Methods section \ref{methods} below) which can achieve low duty cycles of $<$ 1 ppm and low effective sample thermal resistances of $<$ 1 nK-\cmcy /W, and permit measurements at $p \sim 10^{10}$ W/\cmcy. 

When $j$ is increased beyond the central free-flux-flow plateau of \figr{TAFF}, the response once again becomes non-linear and $R$ rises as the pair breaking action of the current suppresses the 
\scv until it is completely quenched and the system enters the normal state when \jj reaches its depairing (pair-breaking) value \jdy . Here, once again, $R$=const. (If $B$=0, the first three distinct stages will be absent and $R$ will rise non-linearly from zero directly to \rrhony .) This \jd is a measure of the superfluid density $\rho_s \equiv 1/\lambda^2$ through the relation \cite{pbreview}:
\be \label{jdbc2} j_{d}(0)=\sqrt{ \frac{8 \Phi_{0} B_{c2}(0) }{ 27 \mu_0^3 \lambda^{4}(0) }} \ene 

\eqr{jdbc2} provides a way to estimate \lam and hence \rrhos through purely transport measurements (as \hcu can also be measured through a transport measurement). This is especially useful when the 
quantity of the \scg material is too small and/or the geometry is unsuitable for susceptibility and inductive type of measurements. Current induced pair breaking also provides another secondary detection of \scvy . Just as a magnetic field shifts a \scg transition downward in $T$, the transition shifts downward with increasing $j$. This current induced shift follows a 
$\Delta T_c \propto j^{2/3}$ behavior. This two-third power law can be used as another confirmation/indication that a resistive 
drop is due to \scv and not some other effect. Like the Meissner effect, \jd goes to the heart of the \scg state through the distinctive way in which a current suppresses \rrhosy . As with the observation of FFF, observing current induced pair breaking requires current densities of the highest magnitude that the \ssc can support. Near the transition, \jd has a smaller value (vanishing as $T \row T_c$); however, the system is especially sensitive to small changes in $T$ that might result from macroscopic Joule heating. Hence this investigation also needs very short low-duty-cycle signals. The next section describes the experimental set up for conducting experiments with high pulsed current densities, and the section that follows summarizes the novel results that have come out of such investigations and how they pertain to the detection and characterization of new near-room-temperature superconducting materials.

\section{Experimental methods}
\label{methods}

\subsection{Electrical measurements} 
\label{measurements} 
Our transport studies in \sscy s include ac, continuous-dc, and pulsed signals. However the most distinguishing capability of our 
group is its expertise in developing techniques and specialized 
measurements using pulsed signals. Fast ($>$100 GHz sampling rate)
digital-storage oscilloscopes (DSO) in conjunction with in-house built pulsed sources and detection electronics are used for this purpose. The specific electronics is customized for every experiment. 

\begin{figure}[h] 
\begin{center} 
 \includegraphics[width=0.5\textwidth]{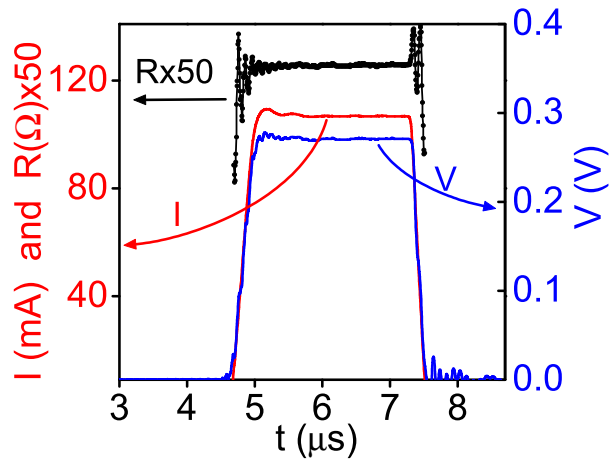}
\end{center} 
\caption{\label{time-response} Some typical quasi-dc high-dissipation pulse waveforms.  
$j = 9.7$ MA/\cmsy, $E = 83$ V/cm, and  $p = jE = 803$ MW/\cmc on the 
plateaus.
(Adapted from Topical Review by M. N. Kunchur, J. Phys.: Condens. Matter {\bf 16}, R1183.)} 
\end{figure}

There are two types of pulsed
measurements. The first type is a quasi-dc measurement, shown in \figr{time-response}, in which \itt and \vtt are are measured on the flat top portions of the pulses corresponding to the zero-frequency limit. The sole purpose of this pulsing is to reduce the duty cycle (\aby 1 ppm) and consequent Joule heating, leading to effective thermal resistances  $R_{th}=\Delta T/dp$ in the 1~nK--1 $\mu$K.cm$^{3}$/W range. 
Pulsing also makes heating at contacts irrelevant because the thermal diffusion distance, 
$\sqrt{D\tau} \sim$ 9--84 $\mu$m,  is shorter than the 
contacts-to-bridge distance (few mm).  

\begin{figure}[h] 
\begin{center} 
 \includegraphics[width=0.4\textwidth]{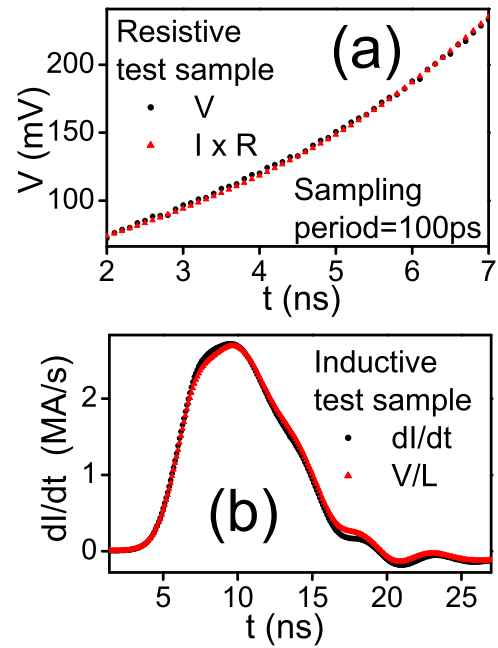} 
\end{center}
\vspace{-2em}
\caption{\label{ballistic-response}{(a) $V(t)$ and $I(t)\times R$ 
for a test resistor and in place of a superconducting sample. (b) 
$dI/dt$ and $V(t)/L$ for a test inductor in place of a superconducting sample.
The voltages across the purely resistive or inductive loads are seen to track
their respective current or current-derivative functions with
subnanosecond accuracy. 
(Adapted from G. F. Saracila and M. N. Kunchur, Phys. Rev. Lett. {\bf 102}, 077001.)}}
\end{figure}

The other type of pulsed measurement explicitly probes the time
dependence of \itt and \vtt and cross correlates them with sub-nanosecond
accuracy. These measurements required significant developments in the sensing circuits for simultaneously \itt and \vtt measurements, and in the preamplifiers 
and pulsed sources. Also an ``artificial ground'' 
approach was developed to better isolate the small signals in the voltage leads  
from the huge fast varying signals in the current leads. 
\figr{ballistic-response} shows the performance of this system under resistive and inductive loads.  
An important distinction of this apparatus is that 
this is not simply an impulse measurement. In measurements 
employing laser pulses or ultrashort current pulses 
generated by Josephson electronics, 
the stimulus is not monitored at all during the measurement and  
only the response is 
monitored. In our setup, both \itt and \vtt are simultaneously 
monitored with high accuracy and can be correlated instant for
instant in the time domain. That makes this technique unique and
opens the doors for correlated \itty--\vtt investigations of other
phenomena in condensed-matter systems.

Further information regarding the
instrumentation and other experimental details can be found  
in two review articles \cite{mplb,pbreview}.

\subsection{Cryogenics}
\label{cryogenics}
The cryostats and magnetic field environments used in our work consist  of 
liquid-helium based as well as closed-cycle systems: (1) An Oxford 
Instruments 16-Tesla superconducting magnet with He$^{4}$ and He$^{3}$ 
cryostats ($0.28 < T < 310$ K). (2) A highly customized Cryomech 
pulsed-tube cryocooler ($2.8 < T < 350$ K) with a 1.2 T rotating water-cooled 
magnet (an advanced triggering method employing acoustic feedback 
 provides extremely stable temperatures for long durations). (3) 
A single-stage ($30 < T < 375$ K) Cryomech pulsed-tube cryocooler that has a very large sample space (\aby150 \cmcy ) and would be especially suitable for measurements on \scg samples maintained in high-pressure cells.  
(4) And a Quantum Design PPMS with a custom insert developed for conducting short-timescale measurements (In order to greatly reduce noise, 
the sample environment and shields of wires within this apparatus are connected to a special highly effective ground consisting of a large loop of 3/4" diameter copper wire buried deep underground. 


\section{Phenomena and regimes that can be studied} 
\label{results}
Our wide-spectrum pulsed-transport measurements enter uncharted space and push 
various parameters to unprecedented extremes (e.g., 
$E > 1000$ V/cm, $p=jE > 10^{10}$ W/cm$^{3}$, and vortex velocity
$v_{\phi} > 10$ km/s). Thus a rich variety of physics is uncovered 
and valuable new information is obtained about fundamental parameters.   
Below we give just a few examples of the types phenomena and regimes that we can study with our techniques. 

\subsection{Ballistic superfluid acceleration and superfluid density suppression}
\begin{figure}[h] 
\begin{center} 
 \includegraphics[width=0.4\textwidth]{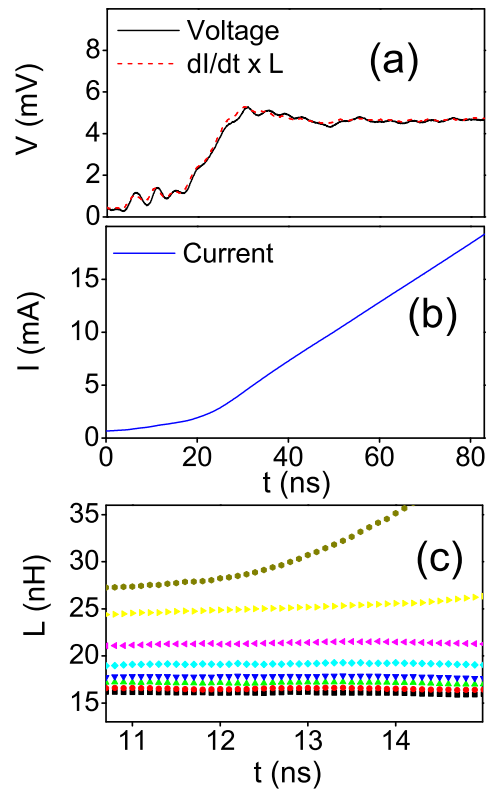} 
\end{center}
\vspace{-2em}
\caption{{(a) Applied voltage plateau \vtt 
for a Niobium sample. (b) Corresponding  \itt function rises 
steadily during the voltage plateau of (a).
The corresponding $dI/dt$ (scaled by a constant $L$) is also plotted on panel (a).
(c) Time dependence of measured total inductance, $L(t) = V(t)/[dI(t)/dt]$
for sample A. The curves are at 
$T$ = 6.10, 5.91, 5.54, 5.17, 4.73, 4.10, and 3.78 K (from top to bottom).
At the highest $T$, $\rho_s$ ($\propto 1/L_k$) is seen to
decline with increasing $j$  ($\propto t$).
(Adapted from G. F. Saracila and M. N. Kunchur, Phys. Rev. Lett. {\bf 102}, 077001.)}}
\label{superacc}
\end{figure}

One of the most direct ways for measuring superfluid density is through its ballistic 
acceleration upon application of an electric field.
From the London equations \cite{tinkhamtext,london}, 
the acceleration of the supercurrent $I_s$ and hence 
$I$ in the external circuit can be written as
\footnote{In more detail \cite{scholten}, 
$L_k \propto I(0,0,T)$, where the superconductor's electromagnetic
response
function $I(\omega,\vec{R},T)$ is defined by $\vec{j}(\vec{r},\omega) =
\frac{e^2N(0)v_F}{2\pi^2 \hbar c} \times \int\frac{\vec{R}
[\vec{R}.\vec{A}_{\omega}(\vec{r'})]}{R^4} I(\omega,\vec{R},T)
d\vec{r'}$,
and $I(\omega,\vec{R},T)$ is related to 
the electron-phonon spectral function $\alpha^2(\omega)
F(\omega)$.}$^{,}$\footnote{The normal-current 
component  $j_{n} \approx (n_{n}/n) \sigma_{n} E$, 
which results from the electric field present during superfluid acceleration, 
is several orders of magnitude smaller than  $j_{s}$ at the frequencies
of the experiment.}:
\be
\frac{dI}{dt} \simeq 
\frac{dI_s}{dt} = \frac{A V\rho_s}{\mu_0 l}
=\frac{V}{L_k}, 
\label{LondonEq}
\ene
where $l$ is the sample length, $A$ is its cross-sectional area, 
and \lk is the kinetic inductance. 
Kinetic inductive effects become prominent close to \tc and in ultranarrow 
interconnects \cite{oppenheim,lee-lemberger}. 
In our work \cite{ballistic}, timescales were chosen to be short enough to have a
sufficient magnitude of $V$ while long enough (compared to 
characteristic timescales such the gap-relaxation and
electron-phonon scattering times) to avoid certain non-equilibrium effects.
Variations in fields occur at length scales that are long compared
with both $\lambda$ and the coherence length $\xi$, so as to avoid
non-local effects. The supercurrent acceleration phase lasts for 
the duration $\Delta t \approx j_c \mu_0 \lambda^2l/V$, where 
\jc 
demarcates the onset of resistance. 
\figr{superacc}(a) and (b) show plots of \vtt and \itt 
for a Nb sample. The \itt function accelerate steadily during the plateau in \vtty . 
The ratio $V(t)/[dI(t)/dt]$  gives the total inductance $L= L_k + L_g$, where $L_g$ is the geometrical
inductance. \figr{superacc}(c) 
shows the time dependence of $L(t)$ at various
temperatures. Besides the increase in the plateau value of $L(t)$ 
with $T$, the curves at highest temperatures show $L(t)$ 
functions that rise with $t$ and hence $I$ (since $I \propto t$ in this interval
as seen in \figr{superacc}(b)). 
 Since $\rho_s \propto 1/L_k$, 
 this is evidence for the suppression of superfluid density 
through the current's pair-breaking action---a regime not seen before 
in any other kind of measurement. 
The fast pulsed signal approach is perhaps 
unique in its ability to probe the current induced 
superfluid suppression regime, since 
continuous-ac probes do not allow high enough excitation levels 
and tunneling measurements reveal the spectral gap $\Omega_g$ rather than \rrhosy . 
This regime is one of the subjects of future study. 
Interesting dependencies in $d \rho_s/dj$ can arise is certain situations---for example 
in the underdoped region of cuprates, where
$j$ may destroy superfluid stiffness without quenching pairing, 
leaving a pseudogap like state of uncondensed pairs above \jd \cite{goren}. 

\subsection{Current induced pair breaking} 
\label{subsecjd}
Current induced pair breaking provides another window to \rrhos as well as the normal state at temperatures below \tcy .
Unlike the ballistic-acceleration measurements which require much faster time resolutions but are conducted in the dissipationless Meissner regime, a \jd measurement is difficult because it enters the dissipative regime and, as mentioned in the Background (section \ref{background}), power densities can get into the GW/\cmc range. Because of these technical difficulties, \jd has been measured in very few \sscy s. Our group was the first to measure \jd in any cuprate \ssc \cite{pair} and in \mgb \cite{mgpair}, and conducted the most comprehensive test of theories over the entire temperature range \cite{pbreview}. 

Fig.~5 
shows an example of current induced pair breaking close to \tcy, which shows 
the classic  $j^{2/3} \propto [T_c-T]$ behavior in the resistive transition shifts. 
Fig.~6 shows an example of high pulsed $IV$ curves, covering the entire temperature range, driving a system into the normal state even at $T<<T_c$ (panel c). 
\begin{figure}\label{slcojd}
 \includegraphics[width=0.4\textwidth]{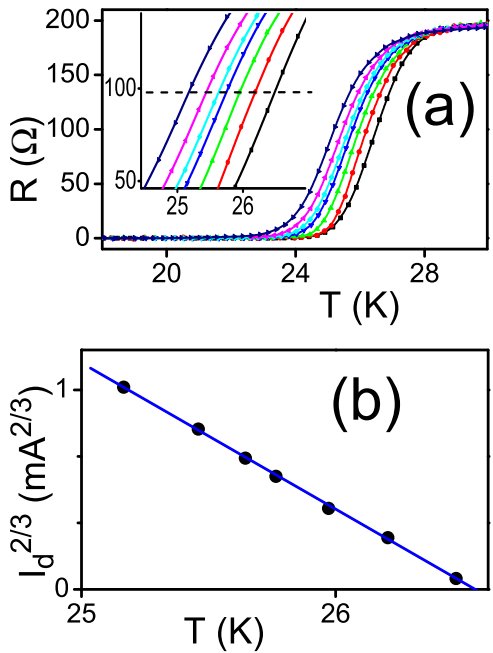} 
\caption{(a) Resistive transitions measured at various values of constant pulsed current in a \slco film sample. (b) $I^{2/3}$ plotted against the $T$ values from the inset of panel (a) where the shifted transition curves 
intersect with the dashed horizontal line.
(Adapted from M. Liang, M. N. Kunchur, L. Fruchter, and Z.Z. Li, Physica C {\bf 492}, 178.)} 
\end{figure}
\begin{figure}\label{mgb2pb}
 \includegraphics[width=0.4\textwidth]{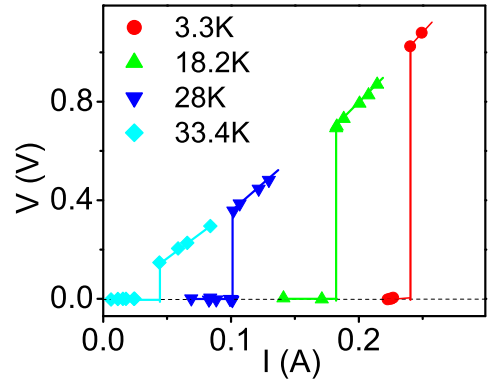} 
\caption{Current-voltage curves for an \mgb sample at various constant temperatures that drive the system into the normal state even at $T<<T_c$. 
(Adapted from Topical Review by M. N. Kunchur, J. Phys.: Condens. Matter {\bf 16}, R1183.)} 
\end{figure}

As discussed in the Background (section \ref{background}), a \jd measurement provides, through \eqr{jdbc2}, a way to obtain \lam and \rrhos through purely transport measurements that is not directly affect by the material's magnetism. We exploited this fact to obtain \rrhos in \nccoy , where a large paramagnetic background due to Nd$^{3+}$ moments precludes an inductive measurement \cite{nccojd}. 

\subsection{Highly driven flux flow}
\subsubsection{Free flux flow: a window to the normal state}

As soon as an applied magnetic field exceeds \hcly , flux vortices penetrate the sample. The Lorentz force of a transport current will then potentially induce flux motion and consequently resistance. Numerous regimes of flux dynamics exist (\cite{blatter} but the simplest is the regime of free flux flow when the pinning force is negligible compared to the Lorentz force. 
However, as explained in the Background (section \ref{background}), the requirement of high driving forces and accompanying dissipation make this FFF regime also very illusive. FFF was quantitatively confirmed for the first time in any superconductor by our group in a \ybco film \cite{fff} (this result is cited in Tinkham's textbook \cite{tinkhamtext}). As explained earlier, the vortex core provides another window to the normal state (which is more easily accessible than destroying the \scv with pair-breaking currents). As shown in 
Fig.~7, 
this concept was fruitfully used \cite{metal} 
to elucidate \rrhony $(T)$ over the entire range for  \ybco and to prove that it had a metallic normal state, i.e., $\rho_n \row$ const as $T \row 0$.
\begin{figure}\label{metalfig}
 \includegraphics[width=0.4\textwidth]{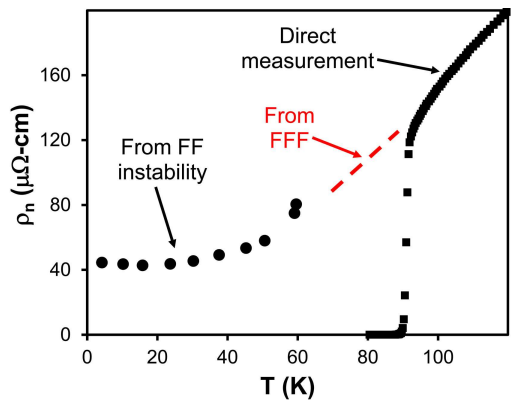} 
\caption{The vortex core as a window to the normal state: Free flux flow 
and the flux-flow instability together elucidate the normal-state resistivity \rrhony $(T)$ in \ybco over the entire temperature range. The measurements indicate a metallic normal state, i.e., $\rho_n \row$ const as $T \row 0$.
(Adapted from M. N. Kunchur, D. K. Christen, and J. M. Phillips, Phys. Rev. Lett. 70, 998; and 
M. N. Kunchur, B. I. Ivlev, D. K. Christen, and J. M. Phillips, Phys. Rev. Lett. 84, 5204.) }
\end{figure}

In other work, combining the extremely low pinning in MoGe alloy films with our high driving forces so that pinning was essentially absent, we were able to sensitively and quantitatively assess the nature of flux dynamics \cite{eval} beyond the Bardeen-Stephen approximation. In particular, we were able to differentiate between the TDGL (time dependent Ginzburg-Landau) theory \cite{ullah,dorsey} and the microscopic LO (Larkin-Ovchinnikov) theory \cite{lo-theory}. As shown in 
Fig.~8, 
we found that the mean-field result arising out of TDGL provides a much better description of FFF than the theory of Larkin and Ovchinnikov; these results represent to most complete tests of the fundamental flux-flow theories \cite{eval}.
\begin{figure} \label{FFFtests}
 \includegraphics[width=0.4\textwidth]{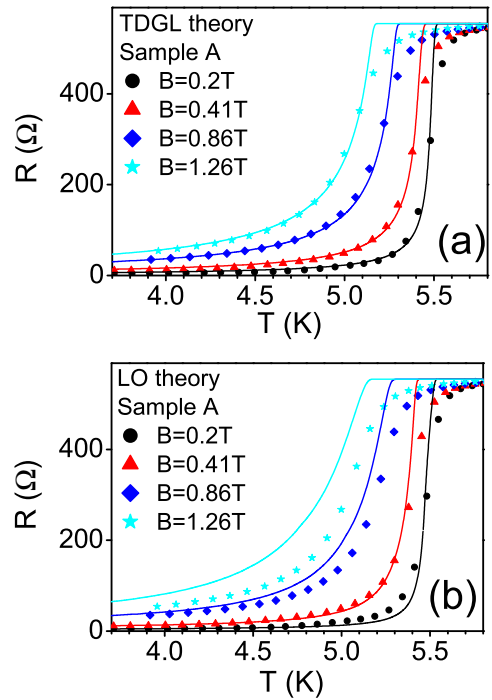} 
\caption{Detailed tests of the ideal free-flux-flow response beyond the Bardeen-Stephen approximation, 
comparing the TDGL and LO theories. These experiments  
provided the first quantitative validation of the TDGL (time dependent Ginzburg-Landau) theory for the FFF response. 
(Adapted from M. Liang, M. N. Kunchur, J. Hua and Z. Xiao, Phys. Rev. B 82, 064502.)} 
\end{figure}

\subsubsection{Flux-flow instability and the spectral function}

\begin{figure}[h]
 \includegraphics[width=2.5in]{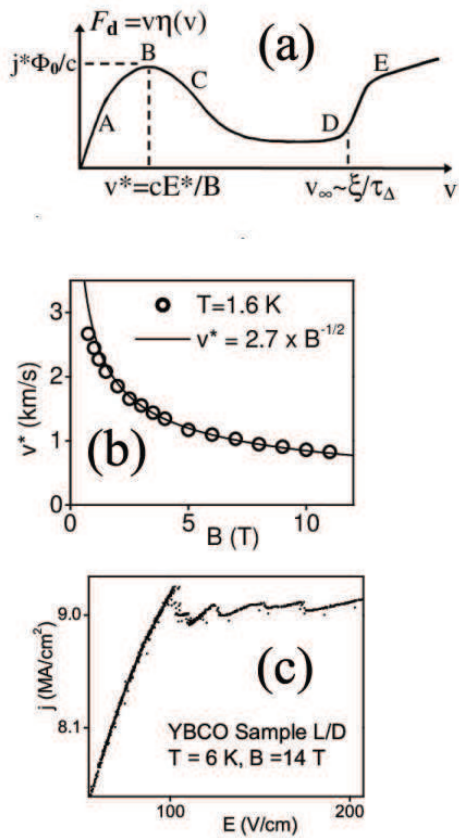} \\
\caption{
(a) The primitive curve 
for the hot-electron regime 
(b) The measured critical vortex velocity \vstary$(B)$ (corresponding to point B on the primitive 
curve)
shows the predicted $1/\sqrt{B}$ dependence.  
(c) Shear distortions cause a sawtooth shaped $IV$ curve beyond the instability.
(Adapted from M. N. Kunchur, B.I. Ivlev,  and J. M. Knight,
Phys. Rev. Lett. {\bf 87}, 177001;  M. N. Kunchur, Phys. Rev. Lett. {\bf 89}, 137005; 
M. N. Kunchur, B.I. Ivlev,  and J. M. Knight, Phys. Rev. B 66, 60505.) }
\label{hot-electron} 
\end{figure}

At high levels of $j$ and $E$, 
the quasiparticle distribution function departs
sufficiently from equilibrium so as to alter
the vortex dynamics. If \rfff rises with $j$ sufficiently rapidly, 
\je turns over and exhibits a negative
differential conductivity as shown in \figr{hot-electron}(a). 
In a current biased measurement this leads to
a flux-flow instability ($E$ jumps above a critical value \estary). 
Two principal scenarios are possible. One occurs
when the average energy increase is not significant but the distribution
function acquires a non-thermal shape---the Larkin-Ovchinnikov (LO) instability 
\cite{lo} 
which occurs close to \tc and when \taueey$\gg$\tauepy . The
other kind is the hot-electron instability, first established by our 
group \cite{unstable,eprelax}, 
which results when the electron temperature rises while
maintaining a thermal like distribution function shape; this instability is
more likely at temperatures well below \tc and when \taueey$\ll$\tauepy.
Flux-flow instabilities have been experimentally investigated for some
time (e.g., Refs.~\cite{musienko}--\cite{grimaldi2010}
but before our group established the hot-electron instability, 
experimental results have been analyzed, sometimes incorrectly, almost entirely in terms of 
the LO effect. Recent works 
have now begun
to interpret their findings using our hot-electron picture. In the
course of investigating the hot-electron instability, we 
discovered new secondary effects (\figr{hot-electron}(b)) related to the fragmentation of the
moving flux into elastic domains, that lead to Gunn-effect like steps in
the $IV$ curve \cite{steps,shear}. \figr{hot-electron} shows our various 
results related to the hot-electron instability.
Ref.~\cite{nori} gives a general overview on the subject of non-monotonic responses and instabilities. 

These flux-flow instabilities serve as tools that yield valuable
information on
the energy relaxation time \taue between quasiparticles and
lattice, which in turn sheds light on $\alpha^{2}F$ 
the electron-phonon spectral 
function\footnote{This spectral function is defined as $\alpha^{2}F =\frac{V}
{(2\pi)^{3}\hbar^{2}} \int
\frac{d^{2}k'}{v'_{F}}|\cal{M}$$_{k=k'}|^{2}$$\delta(\omega -c_{s}|k-k'|)$,
where $\cal{M}$$_{k=k'}$ is the electron-phonon matrix element.}.
It can be shown \cite{eprelax} that $jE \simeq (P_{e} + P_{r})$ in the steady state, where 
the powers  $P_{e}$ and $P_{r}$ correspond to the two lowest order processes
of phonon emission and quasiparticle recombination respectively. 
Each of $P_{e}$ and $P_{r}$ can be calculated in terms of $\alpha^{2}F$; for
example, $P_{e}$ has the form\footnote{Here 
$f_{e}(E,E')=\frac{f(E_{k})[1-f(E'_{k})]}{1 - f(E_{k})} \lb 1 -\frac{\Delta_{k}
\Delta_{k'}}{E_{k}E_{k'}}\rb$ combines the coherence and occupation factors.}
\begin{eqnarray}
P_{e}=\frac{2Vm^{2}}{\pi \hbar^{3}}\int\int\int d\omega d \epsilon d
\epsilon' v_{F} \alpha^{2}(\omega)F(\omega) \hbar \omega \nonumber \\
f_{e}(E,E') 
\delta(E-E' - \hbar \omega) 
\end{eqnarray}
Thus the hot-electron instability gives a weighted average
of $\alpha^{2}F$  that involves purely the processes that transfer energy between 
quasiparticles and phonons. Hence this provides a different window to this
important quantity compared to 	
	some of the traditional techniques such
as tunneling, ARPES, optical conductivity, and neutron scattering. These
other techniques also sense coupling between electrons and other types of excitations
besides just phonons (e.g., optical conductivity also includes the
contribution of spin fluctuations to the inelastic electronic scattering
\cite{carbotte,tu}). 

\section{Concluding remarks} 
\label{conclusions} 
The fast-timescale and high-dissipation techniques that our group has developed over
the past twenty five years provide some unique windows to fundamental parameters
and various physical phenomena and regimes in \sscy s. 

Besides the basic-science importance, these types of measurements also have 
implications for applications, since the experiments chart the detailed growth $E(j)$ 
as the current density $j$ is varied from the onset of dissipation at \jcy , to the complete destruction of superconductivity at the depairing value \jdy . Knowledge of this $E(j)$ function is especially important for applications that operate dissipatively above \jc for short durations 
(such as pulsed superconducting magnets that might be used for 
fusion-energy research) 
but also for applications that operate in the dissipationless mode 
($j$$<$\jcy ) 
to optimize the normally conducting encapsulations of superconductive tapes and wires \cite{apl}.
The fast time-scale studies related to kinetic inductance and superfluid suppression are relevant for devices such as superconducting  photon detectors. 

Thus we believe that our collaborations with groups developing the latest materials, 
in the quest for room-temperature \sscy , will lead to fruitful results. 

\section{Acknowledgements}
\label{acknowledgements}
We would like to acknowledge 
D. K. Christen, C. E. Klabunde, J. M. Phillips, 
D. H. Arcos, C. Wu, G. F. Saracila, M. Liang, N. Shayesteh-Moghadam, S. D. Varner, J. M. Knight, 
B. Ivlev, S. Mejia-Rosales, A. Gurevich,
E.-M. Choi, K. H.P. Kim, W. N. Kang, and S.-I. Lee, 
Y. Cui, A. Pogrebnyakov, P. Orgiani, X. X. Xi, 
P. W. Adams, D. P. Young, 
J. Hua, Z. Xiao, 
A Guarino, A Leo, G Grimaldi, N. Martucciello, C. Dean, M. N. Kunchur, S. Pace, A Nigro,
L. Fruchter, Z.Z. Li,
Q.L. He, H. Liu, J. Wang, R. Lortz, and I.K. Sou. 
This work was supported by the U. S. Department of Energy, Office of Science, Office of Basic Energy Sciences, under grant number DE-FG02-99ER45763.


---------

\end{document}